\documentclass[letterpaper]{article}

\usepackage{natbib,alifeconf}  

\usepackage[colorinlistoftodos]{todonotes}

\usepackage{amsmath,amsfonts,amssymb}
\usepackage{mathtools}
\usepackage{stmaryrd}
\SetSymbolFont{stmry}{bold}{U}{stmry}{m}{n}
\usepackage{bm}
\usepackage{amsthm}
\usepackage{graphicx}
\usepackage{tikz}
\usepackage{tikz-cd}
\usepackage{adjustbox}
\usepackage{quiver}
\usepackage{fontawesome5}
\usepackage[dvipsnames]{xcolor}

\usepackage{url}
\usepackage[hidelinks]{hyperref}
\usepackage{cleveref}
\usepackage{booktabs}

\usepackage{lipsum}

\usepackage{microtype}
\usepackage{silence}
\newtheorem{definition}{Definition} 
\theoremstyle{definition}
\newtheorem{example}{Example}

\newcommand{\mx}{{\mathbf{x}}}
\newcommand{\n}{{\mathsf{n}}}
\newcommand{\mf}{{\mathbf{f}}}
\newcommand{\mzero}{{\mathbf{0}}}

\renewcommand{\mp}{{\mathbf{p}}}
\newcommand{\ms}{{\mathbf{s}}}
\newcommand{\mt}{{\mathbf{t}}}

\newcommand{\mtheta}{{\bm{\theta}}}
\newcommand{\mphi}{{\bm{\phi}}}
\newcommand{\mrs}{{\bm{(r,s)}}}

\newcommand{\ro}{{\text{rocket}}}
\newcommand{\we}{{\text{wedge}}}
\newcommand{\N}{{\mathbb{N}_{\geq 0}}}
\renewcommand{\P}{{\mathcal P}}

\DeclareMathOperator{\M}
{{\mathcal{M}}}

\DeclareMathOperator{\LC}
{{\text{LC}}}

\DeclareMathOperator{\src}
{{src}}
\DeclareMathOperator{\msrc}
{{\mathbf{src}}}
\DeclareMathOperator{\tgt}
{{tgt}}
\DeclareMathOperator{\mtgt}
{{\mathbf{tgt}}}

\newcommand\blfootnote[1]{%
  \begingroup
  \renewcommand\thefootnote{}\footnote{#1}%
  \addtocounter{footnote}{-1}%
  \endgroup
}
\title{Towards chemistries in dynamical systems}

\author{
    Martin Biehl$^{1}$ \and
    Nathaniel Virgo$^{2}$
    \mbox{}\\
    $^1$Cross Labs, Cross Compass, Japan \\
    $^2$University of Hertfordshire, United Kingdom\\
    martin.biehl@cross-compass.com
} 

\begin{document}

\maketitle

\begin{abstract}
Chemistry describes aspects of the universe in terms of molecules and their reactions. In this exploratory work we present a way to describe aspects of any dynamical system in similar terms. 
To describe a dynamical system in this way three decisions have to be made. The first is how many different ``places'' there are at which molecules or chemical species can occur; the second is how to determine the species present (or not) at each  place; and the third is the set of transitions and reactions that can occur between the species in the various places.
For these choices to be compatible with the state update of the dynamical system each state must be able to determine transitions that take the currently occurring molecules to those occurring in the updated state.
We also propose an additional requirement that there is always a unique way to choose the least amount of transitions occurring during state updates. 
We discuss gliders in the game of life cellular and argue that when following their definition of according to Randall Beer they satisfy the additional criterion as well.  
We also point out some issues with the approach. 
\blfootnote{\textcopyright  2026 Martin Biehl, Nathaniel Virgo. Published under a Creative Commons Attribution 4.0 International (CC BY 4.0) license.}
\end{abstract}

\section{Introduction}

In chemistry and also in biology we are often concerned with multiple things (molecules or organisms) of the various kinds or species reacting with each other to produce other such things. We will call such things individualized and refer to them as tokens of various types or species. Evolution, in the way it's usually conceived, also operates on individuals.


In artificial life, cellular automata are often seen as toy universes with simple dynamics and already interesting phenomena. 
Often the focus here is on structures that are also particle-, molecule- or organism like. However, there is no widely accepted formal way to describe these aspects of cellular automata. While this is a main motivation for this work, we take a step further and do not assume a cellular automaton as given but instead only a dynamical system defined on a state set. We believe this will help us better understand what the contribution of the additional structure provided by cellular automata is.   
In a dynamical system, by default, we only have a set of states and a dynamical law and at each time step a single state that the system is in. \emph{A priori} it is neither factorized into configurations of a grid of cells nor composed of tokens of various types. 
We want to understand what is necessary to obtain a chemistry-like description from there. 

Just like actual chemists and biologists aren't looking to describe everything that the laws of physics describe, we are also not necessarily interested in finding descriptions that are equivalent to the original dynamical systems description. We are more interested in describing particular aspects of the dynamical system while ignoring others. A related concept is that of coarse-graining of dynamical systems. Our proposal can be seen as a kind of coarse-graining from a dynamical system to another dynamical system derived from a chemical reaction network or Petri-net. 
In other words, it is a way to partially describe what is happening inside a dynamical system using a chemistry metaphor (i.e.\ Petri-nets).

As mentioned in the abstract this requires three main choices. The choice of an indexed family of functions mapping the state of the system to sets of molecules, the choice of a Petri-net specifying the reactions between the molecules, and the choice of which transitions occur during an update of the systems state.

We note that we only treat finite dynamical systems here. The main reason for this is that the multisets we use are defined to have finite support. So they can't e.g.\ capture an infinite amount of gliders that may exist in the game of life cellular automaton on the standard infinite grid. Extensions to infinite dynamical systems like usual cellular automata on infinite grids might be achieved by patching together finite systems like the ones we consider here. However, the issues we consider here already exist in the finite case and so solving them there first is hopefully already of interest to the reader. 


The rest of the paper is organized as follows. We first provide some required formal definitions and then state the two main formal definitions. The first is the definition of a pattern chemistry and the second contains the additional requirement for what we call an unambiguous pattern chemistry.  Then we explain and discuss the different aspects of these definitions while constructing the pattern chemistry of the glider in the game of life as an example. Finally, we consider some additional examples and provide a short conclusion. 



\subsection{Related work}
The problem of capturing the particle, molecule or organism like structures or patterns in cellular automata has been treated by multiple approaches. The main difference between our and a lot of other work is that we get an explicit interpretation of aspects of the dynamical system or cellular automaton in terms of chemical reaction networks. 
We mention research on coherent patterns and spatiotemporal filtering \citep{shalizi_automatic_2006,lizier2008local,flecker_partial_2011}
which is capable of highlighting patterns of interest in cellular automata but provides no formal link to chemistries. Similarly, \cite{balduzzi_detecting_2011,biehl_towards_2016,krakauer_information_2020,beck_dynamic_2025} propose methods to identify individuals without such a link.
The work most directly related to ours is \citet{beer_investigation_2020} which describes statistical mechanics of gliders using operators to describe creation, annihilation, and persistence. These are similar to the reactions we discuss. The difference is that we eventually want to find out under what conditions some kind of patterns are even worth considering by studying how the underlying system determines these reactions. Therefore we never switch to a statistical description.

\section{Terminology and formal background}
First we define the multisets over a set. A multiset over a set specifies for each element a number of occurrences or tokens of that element. 
\begin{definition}[multiset]
    Given a set $X$ a \emph{multiset of $X$} is a function $\mx:X \to \N$ such that the set of element $x \in X$ with $\mx(x)>0$ (called the \emph{support} of $\mx$) is finite. Accordingly, the \emph{set of multisets of $X$} is defined as
    \begin{align*}
        \M X \coloneqq \{\mx:X \to \N :|\{x \in X:\mx(x)>0\}|\in \N\}. 
    \end{align*}
\end{definition}
   Following \citet{jacobs_multisets_2021} we also write the multiset $\mx$ as a formal sum $\mx=\sum_{x \in X} \mx(x) \,\lvert x\rangle$. This is convenient for calculations.
    We write $\mzero \in \M S$ for the multiset for which, for all $x \in X$ we have $\mzero(x)\coloneqq 0$ and call this the \emph{empty multiset}. 
    We also write $|\mx|\coloneqq \sum_{x \in X} \mx(x)$ for the total number of tokens in the multiset $\mx$.
    
    Note that if we have two multisets $\mx,\bar \mx \in \M X$ we can add them together by adding the numbers for each element:
\begin{align}
    (\mx + \bar \mx)(x) = \mx(x)+\bar \mx(x).
\end{align}
Next we define the powerset.
\begin{definition}
    The \emph{powerset} $\P X$ of a set $X$ is the set of all the subsets of $X$.
\end{definition}
Any (finite) subset $A \in \P X$ can be represented by a multiset $\mx_A \in \M X$ by assigning $1$ to each element of the subset and zero to those elements that aren't in the subset:
\begin{align}
    \mx_A(x)\coloneqq \begin{cases}
        1 \text{ if } x \in A\\
        0 \text{ else.}
    \end{cases}
\end{align}
Many multisets we use in the following are of this kind. 

An important concept here are dynamical systems, so we also define them explicitly. We only consider dynamical systems with finite state spaces in the current work.

\begin{definition}[Dynamical system]
For the purpose of this paper a \emph{dynamical system} is a finite set $X$ called the \emph{state set} together with a function $h:X \to X$ called \emph{update function}.  
\end{definition}
The main example of a dynamical system that we use in this paper is the following. 
\begin{example}[Finite game of life grid]
\label{ex:finitegol}
    The dynamical system $(X,h)$ for a finite grid with game of life dynamics has:
    \begin{itemize}
        \item State set $X\coloneqq2^{\n \times \n}$ where $\n \coloneqq \{1,..,n\}$ for some $n \in \mathbb{N}$. This means $X$ is the set of configurations of an $n$ by $n$ grid of binary cells.
        \item update function $h:X \to X$ given by the rules of the game of life with periodic boundary conditions. 
    \end{itemize}  
\end{example}

The other important concept that is already widely known in the literature are Petri-nets, sometimes also called place-transitions nets.

Petri-nets are a widely used way to describe tokens of various types and their transitions. There can be multiple tokens of each type. Like chemical reactions, these transitions can consume multiple tokens of multiple types and also produce multiple tokens of multiple types. In other words they consume and produce multisets of the set of types. Petri-nets are also known in the chemistry literature as \emph{chemical reaction networks} \citep{feinberg_foundations_2019}, and essentially the same concept appears as the basis of chemical organisation theory \citep{dittrich_chemical_2007}. Apart from this, Petri-nets are also used for to model computation and in particular distributed systems \citep[see e.g.][]{gorrieri_process_2017}. 
Formally they are defined as follows \citep[e.g.\ in][]{meseguer_petri_1990,master_petri_2020}


\begin{definition}
    A \emph{Petri-net} consists of 
    \begin{itemize}
        \item a set $P$ of \emph{places} (also known as species)
        \item a set $T$ of \emph{transitions} (also known as reactions)
        \item a map $\src:T \to \M S$ identifying the multiset $\src(t)$ that transition $t\in T$ consumes
        \item a map $\tgt:T \to \M S$ identifying the multiset $\tgt(t)$ that transition $t \in T$ produces.
    \end{itemize}
\end{definition}
Note that the definition of a Petri-net does not specify which transitions occur when. In this sense they only specify the data necessary to determine what could happen. To describe what can happen we need additional terminology:

\begin{definition}
Assume given a Petri-net $(T,P,\src,\tgt)$ then: 
\begin{itemize}
    \item A \emph{marking of $(T,P,\src,\tgt)$} is a multiset $\mp \in \M P$ of places 
    \item We say a transition $t \in T$ is \emph{enabled in the marking $\mp$} if there exists $\bar \mp \in \M P$ such that $\mp = \bar \mp +\src(t)$.
    %
    \item Given a marking $\mp$ and a  transition $t \in T$ that is enabled, we can remove the source multiset $\src(t)$ from the marking $\mp$ (leaving $\bar\mp$) and add the target multiset: 
    \begin{align}
        \bar \mp + \src(t) \mapsto \bar \mp + \tgt(t)
    \end{align}
    This is referred to as the transition $t$ \emph{firing} and results in the new marking on the right hand side.
 \end{itemize}
\end{definition}

A marking may be thought of as a state of a chemical reactor: for each species it tells us how many molecules of that species are present.
A firing of a transition then represents a reaction occuring, removing its reactants from the reactor and adding its products.


We can also combine multiple transitions in parallel into a so called ``step'' which is a multiset of transitions. This will be used extensively in what follows so we define it separately together with some extra terminology and notation.
\begin{definition}
A \emph{step} is a multiset of transitions $\mt \in \M T$. We can then sum over the source or target sets of the transitions in the step. This results in the \emph{source multiset $\msrc:\M T \to \M S$ of the step $\mt$}: 
\begin{align}
    \msrc(\mt)\coloneqq \sum_{t \in T} \mt(t) \src(t)
\end{align}
and the \emph{target multiset $\mtgt: \M T \to \M S$ of $\mt$}:
\begin{align}
    \mtgt(\mt)\coloneqq \sum_{t \in T} \mt(t) \tgt(t).
\end{align}
A step is \emph{enabled} at a marking $\mp \in \M P$ if all transition tokens in $\mt$ are simultaneously enabled i.e.\ there exists $\bar \mp \in \M P$ such that  
    \begin{align}
        \mp = \bar \mp + \msrc(\mt).
    \end{align}
Similar to a transition, if a step is enabled in marking $\mp$ we can remove the source multisets of all transition tokens in the step and then add all their target multisets
    \begin{align}
        \bar \mp + \msrc(\mt) \mapsto \bar \mp + \mtgt(\mt).  
    \end{align}
In this case we also say that $\mt$ is a step from the marking on the left to the marking on the right. 
\end{definition}
It is important to note that in the following we will only use steps that consume the given marking completely. The standard definition of a step above requires that the source multiset $\msrc(\mt)$ of a step $\mt \in T$ is part of the marking $\mp \in \M P$ by requiring the existence of $\bar \mp \in \M P$ such that  
\begin{align}
    \mp = \bar \mp +  \msrc(\mt).
\end{align}
The idea behind this is that the tokens of $\bar \mp$ can be left untouched by the step and remain available in the marking after the step. So $\bar \mp$ is part of the marking before and after the firing of the step. However, in the case we are interested in, every token in the marking must be consumed by a transition in the step $\mt$ and every token in the marking after the step must be produced by a transition in the step $\mt$. In other words we require $\bar \mp = \mzero$. So a step in our case is only enabled at the marking equal to its source multiset $\src(\mt)$ and always results in the marking equal to its target multiset:
\begin{align}
    \msrc(\mt) \mapsto \mtgt(\mt).
\end{align}

We now give three examples of Petri-nets that are useful to understand the rest of the paper. 

\begin{example}
\label{ex:emptypetri}
Consider the Petri net $(\emptyset,P,\src,\tgt)$ i.e.\ the Petri-net without any transitions. For this Petri-net no marking $\mp$ can be transformed into any other marking $\mp'$, neither by a single transition nor by a step.
\end{example}

\begin{example}
\label{ex:initpetri}
Consider the Petri-net $(\M P \times \M P,P,\src,\tgt)$ with 
\begin{align}
    \src(\mp,\mp')&\coloneqq \mp\\
    \tgt(\mp,\mp')&\coloneqq \mp'.
\end{align}
So this Petri-net consists of one transition for every pair $(\mp,\mp') \in \M P \times \M P$ of markings and that transition consumes the first marking and produces the second. This means that any marking $\ms$ can be transformed into any other marking $\ms'$ using only one transition. So it is never necessary to use a combination of multiple transitions in a step.


\end{example}

\begin{example}
\label{ex:termpetri}
    Consider the Petri-net $(\{\dagger,*\} \times \M P, P, \src,\tgt)$ with 
    \begin{align}
        \src(\dagger,\mp)&\coloneqq \mp\\
        \tgt(\dagger,\mp) &\coloneqq \mzero,
    \end{align}
    and 
    \begin{align}
        \src(*,\mp)&\coloneqq \mzero\\
        \tgt(*,\mp) &\coloneqq \mp.
    \end{align}
    So this Petri-net consists of only of the
    annihilation and creation transitions for all places $p \in P$. 
    Here, we call a transition $t \in T$ the \emph{annihilation transition of $p \in P$} if $\src(t)=\lvert p\rangle$ and $\tgt(t)=\mzero$, so $t=(\dagger,p)$ above. Similarly, a transition $\bar t$ is the \emph{creation transition of $p$} if $\src(\bar t)=\mzero$ and $\tgt(\bar t)=\lvert p\rangle$ so this is $\bar t = (*,p)$ above. 
    
    This Petri-net can also transform any marking $\mp$ into any other marking $\mp'$ by combining the annihilation and creation transitions into a step $\mt(\mp,\mp')$. This step just annihilates all tokens in $\mp$ and creates all tokens in $\mp'$:
\begin{align}
    \label{eq:annihilatecreatestep}
    \mt(\mp,\mp')\coloneqq \sum_{p \in P} \mp(p) \,\lvert \dagger,p\rangle + \sum_{p \in P} \mp'(p) \,\lvert *,p\rangle. 
\end{align}
    Note that the Petri-net from \cref{ex:initpetri} includes all annihilation and creation transitions as well.
\end{example}




\section{Pattern chemistries of a dynamical systems}

In this section we first state the definition of a pattern chemistry and that of an unambiguous pattern chemistry which are the main concepts of this paper. We explain them in some detail afterwards. 

The idea to keep in mind is that a pattern chemistry is a way that an observer tries to partially describe what is happening inside a dynamical system using a chemistry analogy or metaphor (i.e.\ Petri-nets).

\begin{definition}[Pattern chemistry of a dynamical system]
\label{def:patternchem}
Given a dynamical system $(X,h)$ a \emph{pattern chemistry of $(X,h)$} consists of
\begin{enumerate}
    \item an $R$-indexed family of functions $\{\mf_r:X \to \M S\}_{r \in R}$ where $R$ is a finite set called the \emph{set of frames of reference} and $S$ is a finite set called the \emph{set of pattern shapes}
    \item a Petri-net $(T,R \times S,\src,\tgt)$ where $T$ is referred to as the set of \emph{pattern transitions}
    \item a function $\mtheta:X \to \M T$ called the \emph{transition map}.
\end{enumerate}
We then call the function $\mphi:X \to \M (R \times S)$ collecting the results of the family of functions $\{\mf_r\}_{r \in R}$ into 
a single multiset
given by  
\begin{align}
\label{eq:tokenmap}
    \mphi(x) \coloneqq \sum_{r \in R} \sum_{s \in S} \mf_r(x)(s) \,\lvert r,s\rangle
\end{align}
the \emph{token map}. 
The token map $\mphi$, transition map $\mtheta$, and Petri-net $(T,R\times S,\src,\tgt)$ then have to satisfy the following \emph{compatibility condition}: 
For all $x \in X$
\begin{align}
    \mphi(x) &=  \msrc(\mtheta(x)) \label{eq:srccond}\\
    \mphi(h(x)) &=  \mtgt(\mtheta(x)).\label{eq:tgtcond} 
\end{align}
In other words the following diagram commutes:
\[\begin{tikzcd}
	{\mathcal M(R \times S)} && {\mathcal MT} && {\mathcal M (R\times S)} \\
	\\
	X &&&& X
	\arrow["{\mathbf{src}}"', from=1-3, to=1-1]
	\arrow["{\mathbf{tgt}}", from=1-3, to=1-5]
	\arrow["\mphi", from=3-1, to=1-1]
	\arrow["\mtheta"', from=3-1, to=1-3]
	\arrow["h"', from=3-1, to=3-5]
	\arrow["\mphi"', from=3-5, to=1-5]
\end{tikzcd}\]

\end{definition}

\begin{definition}[Unambiguous pattern chemistry of a dynamical system]
\label{def:uapatternchem}
    Given a dynamical system $(X,h)$
    an \emph{unambiguous pattern chemistry} of ($X,h)$ is a pattern chemistry of $(X,h)$ such that, for all $x \in X$ the step $\mphi(x) \in \M T$ returned by the transition map is the unique step minimizing the total number of transitions. That is, for all steps $\mt' \in \M T$ with 
    \begin{align}
    \mphi(x) &=  \msrc(\mt')\\
    \mphi(h(x)) &=  \mtgt(\mt').
\end{align}
    we have $|\mphi(x)|\leq |\mt'|$ and whenever $|\mt'|=|\mphi(x)|$ then $\mt'=\mphi(x)$.
\end{definition}

Over the following few pages we explain the different parts of these definitions and, in parallel, construct an example of a pattern chemistry. The example will use the finite game of life in \cref{ex:finitegol} as the dynamical system and the glider as the described pattern. 

\subsection{The \texorpdfstring{$R$}{R}-indexed family of functions}
\label{sec:rfam}
The rough idea behind the $R$-indexed family of functions $\{\mf_r: X \to \M S\}_{r \in R}$ is that each $\mf_r$ counts the patterns in each of the possible shapes that are found in frame of reference $r \in R$. It tries to capture what it means to view the state of a dynamical system as encoding patterns or objects of various shapes in various places. 

In the example of the finite game of life there are multiple choices. We could take the frames of reference to be the positions $(i,j) \in \n \times \n$, and check at each cell whether there is a glider in one of the possible directions, with one chirality, and in either the rocket or the wedge shape. This would lead to a set $S$ of pattern shapes with $16$ elements. But this suggests a glider has $16$ different shapes when it only really has two, the rocket and the wedge. So we can also define $R$ as the symmetry group $G$ of the finite game of life. Then we can choose the set of pattern shapes as just $S \coloneqq \{\ro,\we\}$. For this we first define the function $\mf_e:X \to \M S$ for the unit element of the symmetry group and then transform it into the function $\mf_g:X \to \M S$ by first acting with the group element on the state $x$ and then applying $\mf_e$. That is, we pick some position, direction, and chirality as the frame of reference $e \in G$, say the position: top left corner $(i,j)=(1,1)$, direction: towards bottom right $d=\searrow$ and chirality: left handed $c=\circlearrowleft$. Then define 
\begin{align}
    \mf_e(x) \coloneqq 
    \begin{cases}
        \lvert\ro\rangle  \text{ if rocket at } ((1,1),\searrow,\circlearrowleft)\\
        \lvert\we\rangle \text{ if wedge at } ((1,1),\searrow,\circlearrowleft)\\
        \mzero \text{ else.}
    \end{cases}
\end{align}
and 
\begin{align}
\label{eq:gliderfam}
    \mf_g(x) \coloneqq \mf_e(g^{-1} \cdot x).
\end{align}
Here $g \cdot x$ refers to the action of the group $G$ on $x \in X$. 
So now we have the $G$-indexed family of functions $\{\mf_g:X \to \M S\}_{g \in G}$ where each $\mf_g$ checks for a different possible position, direction, and chirality given by $g$ whether there is a glider in the rocket shape, or in the wedge shape, or no glider at all.

There are various choices that can be made about what counts as a rocket or wedge being present at a given orientation.
We can, for example, follow the conventions in \citet{beer_cognitive_2014}, in which one layer of boundary cells are included (see figures 2 and 3 in that paper), with the origin point of each pattern in its centre cell.
For $d=\searrow$ and $c=\circlearrowleft$ we then get $\vcenter{\hbox{\includegraphics[width=.1\linewidth]{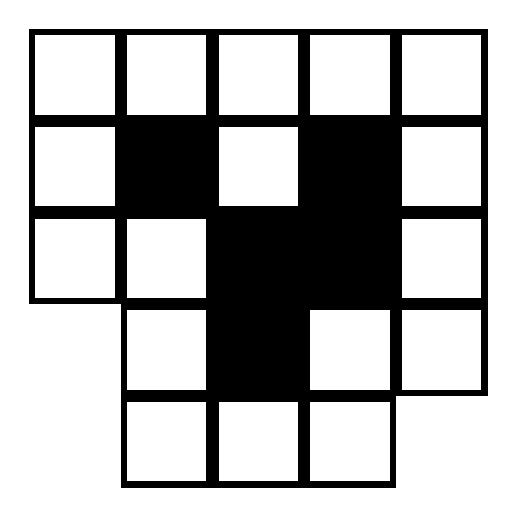}}}$ for the rocket and $\vcenter{\hbox{\includegraphics[width=.1\linewidth]{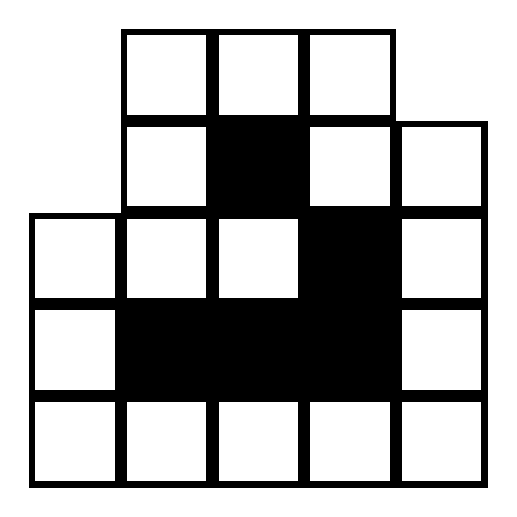}}}$ for the wedge.

Note that the symmetry group of the game of life can be exploited in the way above, but no symmetry group is needed. We could just define each of these functions one after the other by hand. It would only be cumbersome. This is a feature of our approach because we also want to be able to consider patterns in systems that don't have symmetries. 
If we would choose a different dynamics at the edge of the grid instead of periodic boundary conditions, we would lose many of the symmetries of the dynamics but we could still build a similar $R$-indexed family of functions. For the two layers of cells at the edges in particular we would have to choose different functions mapping to $\M S$. The simplest choice would be to map every $x \in X$ to $\mzero$ since no glider-shapes are possible there. Another choice would be to define map other shapes that can occur at the boundary to $\lvert \ro\rangle \in \M S$ or $\lvert \we\rangle \in \M S$. Finally, one could extend our framework to allow a different set $S_r$ of pattern shapes for each frame of reference $r \in R$, but this is beyond the scope of this contribution.

Note that at the same position $(i,j) \in \n \times \n$ there are eight frames of reference, namely those in the four directions and with the two chiralities. However, only one of those eight might have a glider shape present since the presence of one  glider shape in any of those frames of reference rules out the presence of any glider shape in the seven other frames of reference. This need not be the case for all pattern shapes. If we consider a block instead of a glider for example, it would occur in all eight of those frames of reference. So it would produce eight block tokens in total. This highlights a peculiar dependence between the frames of reference and the pattern shapes. One way to try and solve this would be to say that we should indeed only choose one frame of reference for each position. Yet, that would force us to accept that the glider has $16$ shape patterns. We could argue that we should choose the frames of reference according to the symmetries of the shape patterns themselves---but what if we want to track a pattern including both the highly symmetric shapes of the block and the non-symmetric ones of the glider? We would have to choose the frames of reference for the pattern shape with the highest symmetry to prevent one pattern causing multiple tokens. So this could again force us to use a higher number of pattern shapes than we might intuitively want. 
We leave this issue as an open problem. In the examples we are able to choose the $R$-indexed family of function that avoids this issue.

Another thing to note is that the $R$-indexed family of functions is not designed to be a transformation of the whole state of $(X,h)$ into another presentation or coordinate system. It is designed to capture patterns that can be either present or not like gliders in the game of life. In every frame of reference it only distinguishes between states in so far as they produce different pattern shapes and maps all other states to the empty multiset. For example, all configurations of the finite game of life grid in which there is no glider at all have the same value across all frames of reference. 

We chose each $\mf_r$ to return a multiset of $S$. 
Another choice would have been to let $\mf_r$ be a partial function, which would mean it could either return a single pattern shape or ``nothing'', instead of returning a multiset of pattern shapes.
This would prevent two patterns of different shapes to occur in the same frame of reference. However, we would like to allow this since, for example, we could want to track a pattern that can grow in such a way that a small shape $s_1 \in S$ occurs as part a bigger shape $s_2 \in S$ in the same frame of reference. Mapping to the multiset $\M S$ allows setting $\mf_r(x)=|s_1\rangle +|s_2\rangle$, but this isn't possible with a partial function since each $x$ can only be mapped to one element of $S$ (or none).

Mapping to multisets also allows the function $\mf_r$ to return multiple tokens of the same shape. One application is to use a single frame of reference that counts all pattern shapes in the system as in \cref{ex:singleframe} below.

From the $R$-indexed family of functions we calculate the token map, which we discuss next.

\subsection{The Token map}

The token map $\mphi:X \to \M (R \times S)$ is calculated from the $R$-indexed family of functions: 
\begin{align*}
    \mphi(x) \coloneqq \sum_{r \in R} \sum_{s \in S} \mf_r(x)(s) |r,s\rangle.
    \tag{\ref*{eq:tokenmap} repeated}
\end{align*}
In words this says that for each state $x \in X$ the token map returns a multiset over $R \times S$ that is a multiset over the pairs of frames of reference and pattern shapes. Indeed, to each pair $(r,s) \in R \times S$ of a frames of references $r$ and a pattern shape $s$ the token map $\mphi$ assigns the number tokens of that shape $s$ returned by the function $\mf_r$ for that frame of reference $r$. This number of tokens is given by $\mf_r(x)(s)$. 

In the example of gliders in the finite game of life this would return a multiset $\mphi(x) \in \M (G \times S)$ over $G \times S$ and tell us for each frame of reference $g \in G$ where there are rocket or wedge shaped gliders. This will provide for each state $x \in X$ a marking for the Petri-net that is part of the pattern chemistry.

Since $\mphi$ is defined for each state $x \in X$, we can not only consider the marking $\mphi(x)$ of pattern shapes in all frames of reference at $x$ but also the marking $\mphi(h(x))$ of pattern shapes in all frames of reference at the updated state $h(x)$. 

We know that the dynamics $h:X \to X$ of the dynamical system transforms the former into the latter but we want to describe this transformation in a different way: we want to describe it in terms of reactions or transitions of the tokens in the markings. The way to formally specify those reactions or transitions is the Petri-net, which we discuss next.

\subsection{The Petri-net}
The Petri-net $(T,R \times S,\src,\tgt)$ specifies the \emph{possible} transitions of the pattern chemistry. Its set of places $R \times S$ is determined by the $R$-indexed family of functions, but the set of transitions $T$ and the associated source and target functions must be specified additionally. This means we have to specify a set $T$ and for each transition $t \in T$ a source multiset $\src(t) \in \M (R \times S)$ and a target multiset $\tgt(t) \in \M (R \times S)$ over the pairs of frames of reference and pattern shapes.

The idea is that in the end the transition map $\mtheta:X \to \M T$ will determine, depending on the state $x \in X$, which step (multiset of transitions) of the Petri-net is chosen to turn the marking $\phi(x) \in \M (R \times S)$ into the marking $\phi(h(x)) \in \M(R\times S)$. So while the Petri-net defines the transitions that are available to choose from, it is the transition map that is used to pick which transitions \emph{actually} occur. 

In the example of the glider in the finite game of life we would like to choose the Petri-net such that the transitions can transform the gliders in any state $x \in X$ of the grid into the gliders at the state $h(x)$. This will be discussed further below, but it is clear already that the empty Petri-net from \cref{ex:emptypetri} would not be able to do this, while both \cref{ex:initpetri,ex:termpetri} would always be able to this, since they have sufficient transitions to transform any marking into any other.
However, we will argue that choosing different Petri-nets that lie between those examples can be more interesting.

\subsection{The transition map}
\label{sec:transitionmap}
 The transition map $\mtheta:X \to \M T$ assigns to each $x \in X$ a step $\mtheta(x) \in \M T$ i.e.\ a multiset of transitions from the Petri-net of the pattern chemistry. In general it is not determined by the choice of Petri-net although, given a restrictive Petri-net, there might only be one possible transition map. 

 According to the compatibility condition \cref{eq:srccond,eq:tgtcond}, given a Petri net $(T,R \times S,\src,\tgt)$ the transition map has to map each $x \in X$ to a step $\mtheta(x) \in \M T$ that transforms the marking $\mphi(x)$ determined by the token map at $x$ to the marking $\mphi(h(x))$ determined by the token map at the updated state $h(x)$. 

 To see some example transition maps, let us first consider the case where the Petri net is $(\M (R \times S) \times \M (R \times S),R \times S,\src,\tgt)$ and 
\begin{align}
    \src(\mrs,\mrs')&\coloneqq \mrs\\
    \tgt(\mrs,\mrs')&\coloneqq \mrs'.
\end{align}
 as in \cref{ex:initpetri}. For any pair of markings this Petri-net has a single transition transforming the first into the second. So we can satisfy the compatibility conditions simply by defining $\mtheta(x)$ as the step containing the single transition that takes the marking $\mphi(x)$ to $\mphi(h(x))$:
 \begin{align}
 \label{eq:inittmap}
     \mtheta(x)\coloneqq \big\lvert \mphi(x),\mphi(h(x))\big\rangle.
 \end{align}
This gives us a pattern chemistry for every dynamical system, $R$-indexed family of functions and this assumed Petri-net. For each $x \in X$ it uses one transition for the transformation from $\mphi(x)$ to $\mphi(h(x))$.

However, for an observer trying to describe aspects of a dynamical system in terms of molecules and reactions this is unsatisfactory. It ignores that different markings can contain the same tokens and treats each marking as its own indivisible atom that gets transformed into another atom. 

So let us consider the case where the Petri-net is $(\{\dagger,*\} \times \M (R \times S),R \times S,\src,\tgt)$ with 
  \begin{align}
        \src(\dagger,\mrs)&\coloneqq \mrs\\
        \tgt(\dagger,\mrs) &\coloneqq \mzero,
    \end{align}
    and 
    \begin{align}
        \src(*,\mrs)&\coloneqq \mzero\\
        \tgt(*,\mrs) &\coloneqq \mrs.
    \end{align}
as in \cref{ex:termpetri}. This Petri-net also has a step that for any pair of markings transforms the first into the second. It does this by just annihilating all tokens in the first and creating all tokens in the second. For this set
\begin{align}
\label{eq:termtmap}
\begin{split}
    \mtheta(x) \coloneqq& \mt\big(\mphi(x),\mphi(h(x))\big)\\
    \begin{split}
        =& \sum_{(r,s) \in R \times S} \mphi(x)(r,s) \,\lvert\dagger,(r,s)\rangle \\
        &\phantom{\sum}+ \sum_{(r,s) \in R\times S} \mphi(h(x))(r,s) \,\lvert *,(r,s)\rangle.
    \end{split}    
\end{split}    
\end{align}
This also gives us a pattern chemistry for every dynamical system, $R$-indexed family of functions, and this assumed Petri-net. For each $x \in X$ it uses $|\mphi(x)|+|\mphi(h(x))|$ transitions for the transformation from $\mphi(x)$ to $\mphi(h(x))$. Note also that for this Petri-net this is the only choice of the transition function that satisfies the compatibility condition as there is for each pair of markings only one way to build the step that implements the transformation. 

However, this also seems unsatisfactory as a description in terms of molecules and reactions since all reactions are either annihilation or creation. 
We now argue that there are more interesting pattern chemistries using the example of the finite game of life and gliders again.

\subsection{The unambiguous glider chemistry with movement}

Consider the example of the finite game of life grid and the token map 
$\mphi: X \to \M (G \times S)$ derived from the $G$-indexed family of functions given by \cref{eq:gliderfam}. This $\mphi$ produces multisets over $G \times S$ that tell us in which frame of reference $g \in G$ there are gliders and of what shape $s \in \{\ro,\we\}$ in any state $x$ of the grid. As $x$ is updated to $h(x)$ many of those gliders will move in the well known way. If we choose the transition map as in \cref{eq:termtmap} (with $R$ replaced by $G$) then this movement is interpreted as each glider in $\mphi(x)$ getting annihilated and each glider in $\mphi(h(x))$ getting created. If we forget about the group structure of the indices $g$ (as we should here since it could just be a set of indices without group structure), nothing indicates which created glider is a result of the movement of which annihilated glider. In order to introduce connections between these instances of gliders we now introduce additional ``movement'' transitions and redefine the transition map to get a more interesting pattern chemistry. One with glider movement. We will do this next. 

\Citet{beer_bittorio_2020,beer_cognitive_2014} suggested to let the glider transform into any other glider that occurs within its light cone at the next time step. To formulate this here, let $\LC(g,s)$ be the light cone of frame of reference $g$ and pattern shape $s$ and define in addition to existing annihilation and creation transitions $\{\dagger,*\} \times \M(G \times S)$ the set of movement transitions:
\begin{align}
\label{eq:glidermoves}
 \{(g,s)\mapsto(g',s') \in (G \times S)^2: (g',s') \in \LC(g,s)\}  
\end{align}
and
\begin{align}
\src((g,s)\mapsto(g',s'))\coloneqq& \,\lvert g,s\rangle\\  \tgt((g,s)\mapsto(g',s'))\coloneqq& \,\lvert g',s'\rangle.
\end{align}
With this we can indeed construct a more interesting transition map. Using $P=G \times S$ to save space
\begin{align}
\label{eq:glidermovementtmap}      \begin{split}
    &\bar\mtheta(x) \coloneqq \\
        &\sum_{p \in P}\,\sum_{
        p'\in LC(p)} \mphi(x)(p)\mphi(h(x))(p') \,\lvert p\mapsto p'\rangle\\
        &\phantom{=}+ \sum_{p \in P}\,\sum_{
        p'\in LC(p)} \mphi(x)(p)(1-\mphi(h(x))(p')) \,\lvert \dagger,p\rangle\\
        &\phantom{=}+ \sum_{p' \in P}\, \sum_{
        p\in LC(p')}(1-\mphi(x)(p))\mphi(h(x))(p') \,\lvert*,p'\rangle.
    \end{split} 
\end{align}
This expression says to include a movement transition $p \mapsto p'$ whenever the token map assigned a (glider) token at $p$ in state $x$ and at any $p'$ in the light cone of that $p$ in the next state $h(x)$; 
also include an annihilation transition $(\dagger,p)$ whenever the token map assigned a glider token at $p$ in state $x$ and no glider token at any $p'$ in the light cone of $p$ in the next state $h(x)$ (note the use of $(1-\bar\mphi(x))$ for later); 
and finally include a creation transition $(*,p')$ whenever the token map assigned a glider token at $p'$ in state $h(x)$ and no glider token at any $p$ in the light cone of $p'$ in the previous state $x$. 

In \cref{eq:glidermovementtmap} we exploited two things. The first is that there can only ever either be one token at any $p=(g,s)$ by construction of the token map. This allows us to let e.g.\ the product $\mphi(x)(p)(1-\mphi(h(x))(p'))$ compute like a logic gate whether to include an annihilation transition or not. Better notation or an algorithm could also deal with true multisets with possibly multiple tokens in each place. At least as long as the second and more important thing remains the case. 

That is, there is a simple concept (minimizing the transitions used) that made choosing the transitions unambiguous. Note that for each glider movement we could either choose an annihilation transition and a creation transition or a single movement transition. Above we decide to always choose the movement transition whenever we have the option and only if we don't we use annihilation and creation transitions. This also minimizes the total number of transitions used. Importantly, there is no other way to use as few transitions in this case. So minimizing the number of transitions has an unambiguous or unique solution. This makes this glider chemistry an unambiguous pattern chemistry.

To see this note that we never had to make a choice between two possible glider movements where one would have ruled out the other. 
Indeed for each $x \in X$ there cannot be two different possible movement transitions to choose from when transforming the marking $\mphi(x)$ to the marking $\mphi(h(x))$. This would require that there can be: either a single glider in state $x$ and two gliders in state $h(x)$ that both could be the result of the original glider moving; or two gliders in $x$ and a single glider in $h(x)$ that could be the result of either of the two gliders moving. 
Both would require that two different gliders fit into the (future or past) light cone of a single glider. However, it is not hard to see by trying it out that this is impossible.

Of course, these comments are specific to gliders in the game of life; they need not be the case for more general patterns that might be able to reproduce or merge.

Note that, if we consider arbitrary pairs of markings of the Petri-net that includes the annihilation, creation, and movement transitions and not just those that actually occur as results of applying the token map to states $x,h(x) \in X$ then there can be ambiguities. 
The set of multisets $\M (G \times S)$ over frames of references and shapes contains multisets like $|g',s'\rangle +|g'',s''\rangle$ with two different glider tokens in the light cone of a third $|g,s\rangle$. To transform the latter multiset into the former while minimizing the number of transitions we would not know whether to choose the transition $(g,s) \mapsto (g',s')$ and a creation transition $(*,(g'',s''))$ or the transition $(g,s) \mapsto (g'',s'')$ and a creation transition $(*,(g',s'))$. However, these multisets never actually occur as the result of the token map $\theta:X \to \M (G \times S)$ acting on a state of the game of life grid because the dynamical system $(X,h)$ doesn't allow them.      

\subsection{The unambiguous pattern chemistries}
The condition for unambiguous pattern chemistries requires that for all pairs of markings $(\mphi(x),\mphi(h(x))) \in \M (R \times S)^2$ that actually occur as a result of an update from a state $x$ to a state $h(x)$ in the dynamical system there is only one way to use the least possible amount of transitions. 

From the point of view of an observer trying to describe the dynamical system $(X,h)$ in terms of something like molecules and reactions this seems a reasonable desideratum. Minimizing the number of transitions simplifies the description of what is happening during an update step. Ambiguity about how to minimize this number would require the observer to make further choices about how to describe what happens during such an update. 

A more ambitious and currently speculative claim is that it is the non-ambiguity of the glider's movement transitions that convinces us that it actually is a moving object and that this generalizes to other patterns and other transition types like replication and merging.

Concerning the more practical matter of finding interesting pattern chemistries we might always start from the pattern chemistry with only annihilation and creation transitions. Then we can try to add further transitions like those for movement but also possibly for replication, merging, and ``multiple input-multiple output'' transitions while making sure they don't introduce ambiguities. 

\subsection{Other examples}
\begin{example}[On cells in finite game of life]
    Consider again the finite game of life as the dynamical system $(X,h)$. Instead of gliders we here want to consider just the ``alive'' or ``1'' cells as the objects or molecules. For this we define a $R$-indexed family of functions with $R= \n \times \n$ indicating the cell position. The set of pattern shapes is the singleton state $S=\{1\}$. We then choose 
\begin{align}
    \mf_{i,j}(x)\coloneqq \begin{cases}
        |1\rangle &\text{ if } x_{i,j}=1\\
        \mzero &\text{ else.}
    \end{cases}
\end{align} 
Then we find that $\mtheta:X \to \M (\n \times \n)$ just returns the multiset with one token at each index whose cell is on:  
\begin{align}
    \mphi(x) =&\sum_{(i,j)\, \in\, \{(k,l) \,\in\, \n \times \n\,:\,x(k,l)\,=\,1\}} |i,j\rangle.
\end{align}
We then can define a Petri-net with the set of places $P=\n \times \n$ and the usual annihilation and creation transitions. If we want to define movement based on the light cone for these on patterns we immediately introduce ambiguities. Clearly, we can have on cells within the light cones of other on cells, as by definition three on cells in the light cone of any cell cause an on cell at the next time step. So such light cone movement transitions are ambiguous. We could introduce a movement transition into a single direction but this would be like shifting the entire grid in the opposite direction at each time step and not a convincing notion of movement. The more natural choice would be to only allow ``movement'' in place, i.e.\ more like persistence. This would remain unambiguous. 

Note that when talking about the game of life people rarely refer to the on cells as moving but do often refer to them as persisting. This lends a bit more credibility to the hypothesis that non-ambiguous transitions capture some intuitive human attributions to the patterns in cellular automata.
\end{example}

\begin{example}
\label{ex:singleframe}
    Now consider the finite game of life grid again but this time let there only be a single frame of reference $R=\{\bullet\}$. Let $S = \{\ro,\we\}$ and \begin{align}
        f_\bullet\coloneqq  \sum_{s \in S} \sum_{g \in G}\mf_g(x)(s) \,\lvert s\rangle.
    \end{align}
    This counts all rocket and wedge shaped gliders in the entire grid and keeps count of each shape separately. So this example uses multisets that are not just subsets in disguise. Since $R=\{\bullet\}$  the above $\mf_\bullet$ is also the token map here: 
    $\mphi=\mf_\bullet$. 
    If we choose the Petri-net $((\dagger,*) \times \M S,S,\src,\tgt)$ with only annihilation and creation transitions then the standard transition map \cref{eq:annihilatecreatestep}works in this case as well. If we want to introduce the glider movement transitions from \cref{eq:glidermoves} we see that they require source and target multisets over $G \times S$ and not just over $S$. As this set only has two elements we can only define the $4$ movement transitions $\{s\mapsto s'\in S^2\}$ between the $2$ shapes. To get a transition map $\mtheta:X \to \M T$ note it has access to the whole grid state. It can therefore, in exactly the same way as the transition map in \cref{eq:glidermovementtmap}, identify all movement and other transitions. Then it can just count how many there are of each of the three available types: $s \to s', (\dagger,s)$, and $(*,s')$. So we use the definition of $\bar \mtheta$ from \cref{eq:glidermovementtmap} to compute a transition map for this case here (also turn $P$ back into $G\times S$):
    \begin{align}
    \begin{split}
    \bar{\bar \mtheta}(x)\coloneqq& \sum_{(g,g') \in G^2} \bar\mtheta(x)((g,s) \mapsto (g',s'))|s \mapsto s'\rangle \\
    &+ \sum_{g \in G} \bar \mtheta(x)(\dagger,g,s)|\dagger,s\rangle \\
    &+ \sum_{g \in G} \bar \mtheta(x)(*,g,s)|*,s\rangle.
    \end{split}
\end{align}
    This satisfies the compatibility conditions and so gives us a pattern chemistry. It is not unambiguous. If we have, in the step $\bar{\bar \mtheta}(x)$, for example one transition $(\ro \mapsto \we)$ and one $(\we \mapsto \ro)$ then we can always replace them with one transition $(\ro \mapsto \ro)$ and one $(\we \mapsto \we)$. The new and different step still transforms the same marking into the same marking and has the same amount of transitions so the ones with the least amount of transitions aren't unique.

\end{example}

\section{Conclusion}
We have proposed a definition of ``pattern chemistries'' which provide a formal way to describe aspects of a dynamical system in terms of tokens of various types and reactions between them. We also proposed a criterion requiring such pattern chemistries to have an unambiguous way to choose the least amount of transitions occurring during each update of the dynamical system. We discussed the definition and some examples, in particular the glider in the game of life. As mentioned, an open problem is how to make sure the choice of the $R$-indexed family of functions is not degenerate. More work is also needed to better understand the criterion for interesting pattern chemistries and derive practical ways to find them. Finally, we would like to establish the relation to existing work on the glider in the game of life in \citet{beer_cognitive_2014,beer_bittorio_2020} and also the theoretical physics account of particles in field theories.

\section{Acknowledgements}
MB and NV are funded by the Advanced Research + Invention Agency (ARIA) through project code MSAI-SE01-P011. MB's work on this was also made possible through the support of
Grant 62828 from the John Templeton Foundation. The
opinions expressed in this publication are those of the authors and do not necessarily reflect the views of the John
Templeton Foundation.

\footnotesize
\bibliographystyle{apalike}
\bibliography{bibliography} 

\end{document}